# CHMMOTv1 - Cardiac and Hepatic Multi-Echo (T2$^*$) MRI Images and Clinical Dataset for Iron Overload on Thalassemia Patients

Iraj Abedi[1], Maryam Zamanian[1], Hamidreza Bolhasani[2], Milad Jalilian[3*]

[1] Department of Medical Physics, School of Medicine, Isfahan University of Medical Sciences, Isfahan, Iran.
[2] Department of Computer Engineering, Islamic Azad University Science and Research Branch, Tehran, Iran.
[3] Department of Neuroscience, Neuroimaging and Addiction Studies, School of Advanced Technologies in Medicine, Tehran University of Medical Sciences, Tehran, Iran.

**\***Corresponding author
mjalilian@razi.tums.ac.ir

**Abstract**

Owing to the invasiveness and low accuracy of other tests, including biopsy and ferritin levels, magnetic resonance imaging (T2 and T2*-MRI) has been considered the standard test for patients with thalassemia (THM). Regarding deep learning networks in medical sciences for improving diagnosis and treatment purposes and the existence of minimal resources for them, we decided to provide a set of magnetic resonance images of the cardiac and hepatic organs. The dataset included 124 patients (67 women and 57 men) with a THM age range of (5-52) years. In addition, patients were divided into two groups: with follow-up (1-5 times) at time intervals of about (5-6) months and without follow-up. Also, T2* and, R2* values, the results of the cardiac and hepatic report (normal, mild, moderate, severe, and very severe), and laboratory tests including Ferritin, Bilirubin (D, and T), AST, ALT, and ALP levels were provided as an Excel file. This dataset CHMMOTv1) has been published in Mendeley Dataverse and is accessible through the web at: http://databiox.com.

**Abbreviations**

IrO: Iron overload; α-THM: α-thalassemia; β-THM: β- thalassemia; PHT: pulmonary hypertension; MRI: magnetic resonance imaging; CMR: cardiac MR; HMR: hepatic MR.

## 1. Introduction

Thalassemia (THM) is one of the most inherited hemoglobinopathies and the most common monogenic disorder worldwide. Every year, approximately 300-400 thousand fetal-affected types with anemia are born (1, 2). Due to the location of the mutation in the hemoglobin chain, it is classified into two groups: α-thalassemia (α-THM) as the fetal type and β-thalassemia (β-THM) divided into minor and major types (1, 3). The need for frequent blood transfusions to relieve the effects of anemia can lead to complications, such as iron overload (IrO) in the cardiac, hepatic, and endocrine glands. Regular chelation therapy and early diagnosis are important for dangerous complications of iron in the liver such as hepatomegaly (fibrosis and cirrhosis), cardiac failure, systolic and/or diastolic left ventricle (LV) dysfunction, pulmonary hypertension (PHT), and arrhythmia leading to death (4).

Methods include a hepatic biopsy, serum iron and ferritin levels, transferrin saturation, and magnetic resonance imaging (MRI) were used to calculate IrO. Although serum ferritin level estimation is inexpensive and the most accessible method for assessing the body's iron concentration because it shows the short-term total iron of the body, this test has low precision (5, 6). Non-transferrin-bound iron and redox-fraction-sensitive plasma iron are also very complex, and because their results are affected by the last treatment injection, they are less valid than other tests (7). The biopsy method is considered the "Gold standard," but is an invasive and dangerous method, especially for the cardiac, and has serious errors in evaluation (8, 9). Therefore, MRI-T2*, is considered a reliable and standard method for evaluating IrO in cardiac and hepatic organs (10, 11).

MRI was first used in patients with THM in 2001 by Anderson et al., and its use has increased rapidly since then. These images show the indirect effects of iron on local hydrogen protons (12). The main techniques used to measure IrO are the signal intensity ratio and the T2 and T2* times. Relaxometry methods including T2 and T2* have been considered because of the limitations of the signal-attribution technique (13, 14). In these methods, hepatic iron levels are obtained using different echo times (TEs) during breath-holding with increasing TE; the faster the curve decreases (the lower the T2 and T2* values), the more iron there is in the tissue, and the darker the image (15, 16). Also, studies have shown that T2* provides the most excellent estimates up to a 24-week interval (17, 18). Another method is the rrelaxometry technique using a single multi-echo gradient-echo T2 sequence which can be expressed as relativity rates: R2 (1/T2) or R2* (1/T2*)(19).

The ability of artificial intelligence to facilitate diagnosis and treatment procedures and the need for data related to the same domain for training neural networks highlight the importance of preparing datasets with the aim of speeding up the work. This study is the first to simultaneously prepare a set of cardiac MR images (CMR) and hepatic MR images (HMR) for the analysis and image processing of THM patients.

## 2. Related Works

To date, the only dataset provider study for patients with THM was conducted by Shiae et al. in Mashhad, Iran, between February 2016 and January 2019, in the form of open-source CMR images of 50 subjects, including 37 THM patients and 13 healthy subjects, with clinical and echocardiographic data, such as clinical signs of heart failure, shortness of breath, decreased activity, hand and foot swelling, round the eye and chest pain, and arrhythmia. All images were

16-bit grayscale with a resolution of (192 × 256) pixels, stored in DICOM format, and finally compressed and saved in RAR format (20).

## 3. Data Acquisition and Analysis

All T2* MR images were obtained using GE Healthcare (Waukesha, USA). Cardiac gating MRI examination was performed using a single mid-papillary ventricular short-axis slice in the supine position with a single breath-hold using a torso phased-array body coil, and a multi-echo gradient-echo sequence (12 echoes) was obtained. The field of view (FOV) in CMR extends caudally from the carina to the lower renal pole (40 × 40) cm, matrix size (128 × 116), slice thickness 10 mm, 12 different TEs (1.8, and 17.9) ms, TR (31.3) ms, and bandwidth 1562 Bw/pixel. An execution time of more than 20 ms was considered a normal IrO, between 15 and 20 ms as mild, (10 –15) ms as moderate, and less than 10 ms as indicative of severe myocardial siderosis.

For the HMR, a single breath-hold technique using a multi-echo gradient-echo in 12 different TEs (1, 15.1) ms, FOV (40 × 40) cm, matrix size (128 × 116), voxel size should have 3.1 * 3.4 mm in-plane resolution, slice thickness 8 mm, TR 120 ms, and bandwidth 1736 BW/pixel. The flip angles for both CMR and HMR were $20^0$. T2* values of less than 30 ms indicate hepatic abnormal IrO: mild (>6/2) ms, moderate (3.1-5.2) ms, severe (2.1-3. 1) ms, and very severe (<2.1) ms.

## 4. T2*- MRI Database Specification

The current dataset included 210 MRI image files (every file different series) of 124 patients with THM, including 67 women and 57 men age range (5-52) years. The data were divided into two groups with 75 patients with follow-up (Between 1-5 times) in time intervals of about (5-6) months and 48 patients without follow-up. The image format was DICOM, with a 16-bit grayscale resolution of (192 × 256) pixels. After anonymization to protect the security of the patients, the image files were stored in RAR files.

In addition, an Excel file containing T2* and R2* values and a report on cardiac and hepatic IrO of the patients (normal, mild, moderate, severe, and very severe) was provided. Figure 1 shows a sample of images from the current dataset.

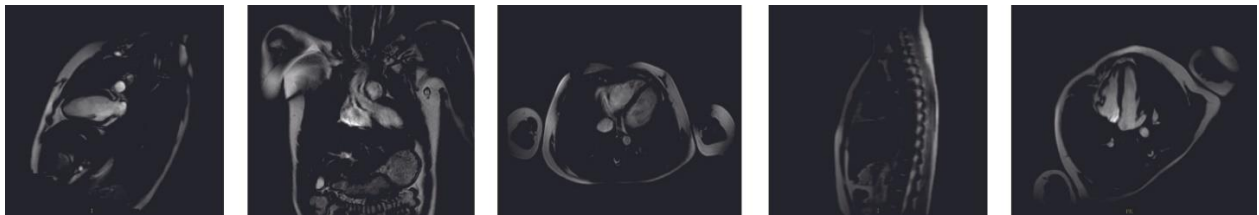

Figure 1. A part of the MRI image is related to a thalassemia (THM) patient of different sections whose iron overload (IrO) condition of the heart is severe and very severe in the liver.

The sex separation of the patients as well as their cardiac and hepatic IrO conditions are shown in detail in Figures 2 and 3. The patient's blood test results, including Ferritin, Bilirubin (D and T), aspartate aminotransferase (AST), alanine aminotransferase (ALT), and alkaline phosphatase (ALP) levels, were provided in an Excel file.

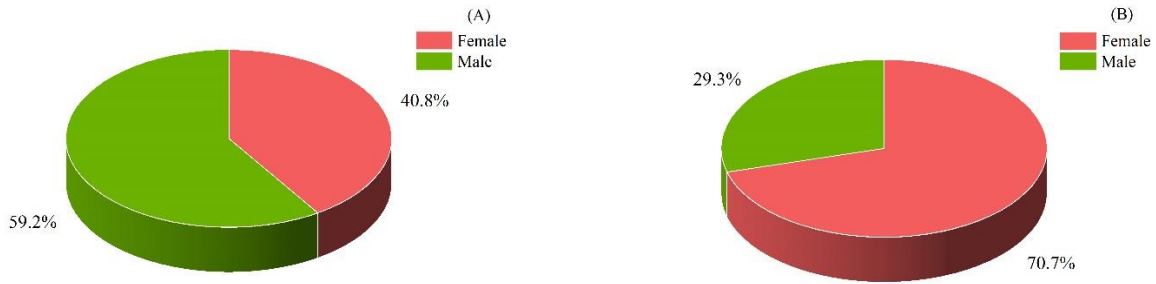

Figure 2. Distribution of the number of men and women in the two categories with (A) and without follow-up (B)

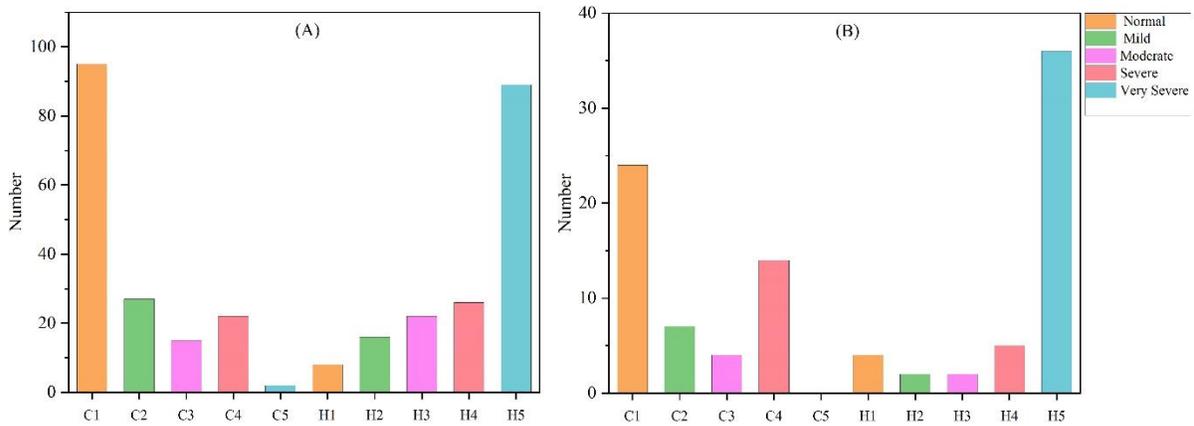

Figure 3. In part A, the heart and hepatic situation of patients with follow-up and in part B of patients without follow-up (C: Cardiac, H: Hepatic, and (1,2,3,4, and 5) number are (Normal, Mild, Moderate, Severe, and Very Severe, respectively).

## 6. Authors Biography

**Iraj Abedi, PhD**

Assistant Professor,

Medical Physics Group, School of Medicine,

Isfahan University of Medical Sciences, Isfahan, Iran

Fields of Interest: Medical Physics, Radiation Therapy and Artificial Intelligence in Medical Sciences.

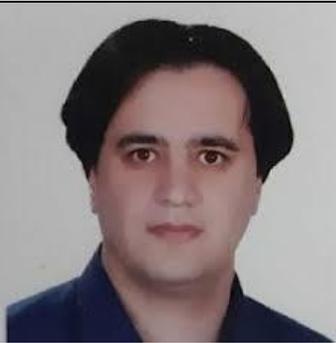

**Maryam Zamanian, MSc**

BSc in Nuclear Medicine at Kermanshah University of Medical Sciences (KUMS) and MSc of Medical Physics from Isfahan University of Medical Sciences (IUMS).

Fields of Interest: Radiology [Nuclear Medicine, and Magnetic Resonance Imaging], Radiation Therapy, Artificial Intelligence in Medical Sciences.

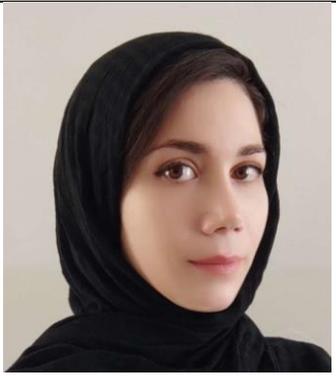

**Hamidreza Bolhasani, PhD**

AI/ML Researcher / Visiting Professor

Founder and Chief Data Scientist at DataBiox

Ph.D. Computer Engineering from Science and Research Branch, Islamic Azad University, Tehran, Iran. 2018-2023.

Fields of Interest: Machine Learning, Deep Learning, Neural Networks, Computer Architecture, Bioinformatics

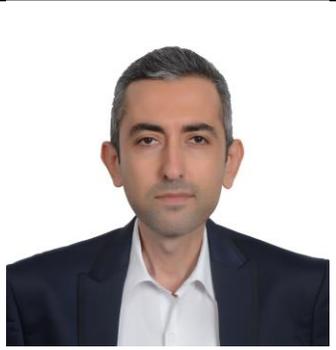

**Milad Jalilian, PhD**

BSc in Radiology Technologist at Behbahan University of Medial Sciences. MSc in Medical Imaging Technology at Isfahan university of Medical Sciences.

PhD Student of Neuroimaging at Tehran University of Medical Sciences (TUMS).

Fields of Interest: Neuroimaging, Biomedical Engineering, Image and Signal Processing.

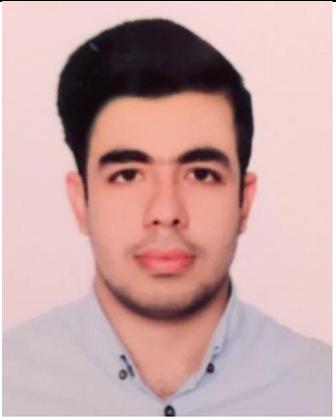